\documentstyle[aps]{revtex}

\begin{document}

\title{Quantal distribution functions in non-extensive statistics
and an early universe test revisited}

\author{U\v{g}ur T{\i}rnakl{\i}$^{1,}$\thanks{tirnakli@sci.ege.edu.tr}
and
Diego F. Torres$^{2,}$\thanks{dtorres@venus.fisica.unlp.edu.ar}}

\address{$^1$ Department of Physics, Faculty of Science, Ege University
35100 Izmir-Turkey\\
$^2$ Departamento de F\'{\i}sica, 
Universidad Nacional de La Plata,
C.C. 67, 1900, La Plata,  Argentina}

\maketitle

\begin{abstract}
Within the context of non-extensive thermostatistics,
we use  the 
factorization approximation to study 
a recently proposed early universe test. 
A very restrictive bound upon the non-extensive
parameter is presented: $|q-1| < 4.01 \times 10^{-3}$. \\

\noindent 
{\it PACS Number(s):} 05.20.-y, 05.30.-d, 05.70.-a, 26.35.+c, 98.80.Ft

\end{abstract}

\vspace{1.5cm}

Although nonextensive formalisms are much in vogue in Physics, a
very interesting one, the so-called Tsallis Thermostatistics (TT), seems to
be one of the most actively studied.   The
proposed formalism basically relies upon 
two postulates \cite{first,second}:
\begin{itemize}
\item A new definition 
of a nonextensive entropy,  $$S_q=k(1-\sum_i^W p_i^q)/(q-1) .$$

\item   A new definition of expectation value, 
$$\left<{\cal O}\right>_q=\sum_i p_i^q{\cal O}_i .$$
\end{itemize}
TT introduces a new parameter, $q\in\Re$, which is usually called the
nonextensivity parameter or the Tsallis $q$-index, and it
contains the standard, extensive Boltzmann-Gibbs statistics, as a special
case where $q$ is taken to be unity.
Up to present days, TT has been found to 
admit generalizations of some of the important
concepts of statistical physics \cite{concepts},  and  to yield results which can
explain some observational and experimental data where standard
statistics is known to fail \cite{verifications}.
All these efforts accelerate new attempts to find the physical meaning of
the nonextensivity parameter, a long standing puzzle which has only now
started to be clarified. 
On one side, some works have been devoted to the study of dynamical
and dissipative systems \cite{fractality}. On the other, an entirely new fractal
canonical ensemble was introduced in order to relate TT 
with a scale invariant thermodynamics \cite{ALEMANY}. 
Of course, an alternative
way to search for the meaning of $q$ is 
related with the estimation of bounds in 
measurable physical systems. We mention 
the study of the microwave
background radiation \cite{tsallis,tirnakli1}, the Stefan-Boltzmann
constant \cite{plastino,tirnakli2}, 
the early Universe \cite{torres1,torres3} and the
primordial neutron to baryon ratio in a cosmological expanding
background \cite{torres2}. 
In these applications, except in Refs. \cite{tirnakli1,tirnakli2},
the quantal distribution functions of TT were obtained
by an asymptotic approach of the 
kind $\beta (1-q) \rightarrow 0$, where $\beta$
is the usual inverse temperature. However, quantal distribution functions have 
been previously generalized by B\"{u}y\"{u}kk{\i}l{\i}\c{c} et al. 
\cite{buyukkilic}
using a rather straightforward procedure, referred to 
as Factorization Approximation.
The formulae so obtained have proved to be simpler and more general than the
ones derived with the common Tsallis et al. method \cite{tsallis}. Simpler,
because the steps followed in the derivation are completely the same when 
compared with standard textbooks. More general, because the factorization
approach does not need a value of $q \simeq 1$.

Quantal distribution functions within the
factorization approximation were, however, regarded as a rather rough
technique because of a work by Pennini et al. \cite{pennini}. They
considered fermion and boson systems with very small occupation numbers.
However, very recently, Wang and L\'e M\'ehaut\'e 
\cite{wang} analysed the problem in detail and showed
that there exist a temperature interval, a 
{\it forbidden zone}, where the
deviation from the exact result  is significant, but  
that outside this
zone, the factorization approach results could be used 
with confidence. In addition, they
verified that the magnitude of the {\it forbidden zone} remained constant
with the increase of the number of particles,  contrary
to the result stated in Ref.~\cite{pennini}. This fact might 
motivate new efforts for the study of macroscopic systems (where the number
of particles is $\sim 10^{23}$) within the simpler approach.
The generalized distribution functions of the factorization approach 
could be used at temperatures up to about $10^{20}$ K for such a 
system \cite{wang}.

All the above remarks led to recompute some bounds already 
obtained within the 
Tsallis et al. approach in this simpler framework,
in order to get more reliable results 
and to check for consistency. This is what we briefly do below concerning the
early universe test proposed in \cite{torres1}.

The test devised consist in compute a first order deviation in $(q-1)$ to the
temperature of freezing out of the weak interctions in the early universe, $T_f$.
This temperature is essential in the primordial nucleosynthesis scenario which,
basically,  is the competition between the rate of the expansion
of the universe and that of the weak interactions which regulates
the conversion of neutrons into protons and viceversa. 
When the expansion exceeds
the rate of interactions they freeze out, and the final yields for the element
production are roughly 
the ones we observe today. In particular, how a deviation in
$T_f$ cause a different prediction for the helium primordial yield, $Y_p$,
was made by Casas et. al. \cite{CASAS}. The result is,

\begin{equation}
\delta Y_p=Y_p\left[ \left( 1-\frac{Y_p}{2\lambda }\right) \ln \left( 
\frac{2\lambda }{Y_p}-1\right) +\frac{-2 t_f}{\tau _n} \right]
\frac{\delta T_f}{T_f}.
\end{equation}
Here, a 
radiation era relationship between time and temperature of the form 
$(T \propto t^{-\frac{1}{2}})$ is assumed \cite{BARRACO} and   
one sets $\delta T_{nuc}=0$, because it is fixed by the binding 
energy of the deuteron. 
$\lambda = \exp (-(t_{nuc}-t_{f})/\tau )$ 
stands for the fraction of
neutrons which decayed into protons between $t_{f}$ and $t_{nuc}$, 
with $t_{f}$ $(t_{nuc})$ the time of freeze out of the weak 
interactions (nucleosynthesis) and $\tau$
the neutron mean lifetime.
Considering now, conservatively, 
$Y_p=Y_p^{obs}=0.23$ and $|\delta Y_p |=0.01$, which is the observational
error for $Y_p$, and standard values for the times and the mean life
of neutron --which in fact, is not modified at order $(q-1)$--, we
must ask for 

\begin{equation}
\label{b2}
0.01 > 0.3766 |\frac{\delta T_f}{T_f}| .
\end{equation} 
A more detailed account of the processes that occurs 
in the early universe within this statistical framework 
is given elsewhere \cite{torres2}.

To compute  the  $(q-1)$ corrections to $T_f$ we recall the output of the
factorization approach. The quantal distribution functions are given by,

\begin{equation}
\label{1}
n_{q\,[bosons]} = \frac {1 }{e^x - 1} - \frac {(1-q)}{2} \frac {x^2 e^x}{(e^x -1)^2},
\end{equation}
\begin{equation}
\label{2}
n_{q\,[fermions]} = \frac {1 }{e^x + 1} - \frac {(1-q)}{2} \frac {x^2 e^x}{(e^x + 1)^2}.
\end{equation}
There  are two main corrections acting upon $T_f$. The first comes from 
the computation of the energy density of the universe. 
When the particles are higly
relativistic, $T \gg m$, and non-degenerate $T \gg \mu$, we get

\begin{equation}
\rho_{bosons} = \frac{g_b}{2 \pi^2} \int_0^\infty dE E^3  n_{q\,[bosons]},
\end{equation}
\begin{equation}
\rho_{fermions} = \frac{g_f}{2 \pi^2} \int_0^\infty dE E^3  n_{q\,[fermions]},
\end{equation}
$g_{b,f}$ stands for the degeneracy factor of each one of the species
involved. Using (\ref{1},\ref{2}), we finally obtain

\begin{equation}
\label{rho}
\rho_{total}=
\rho_{bosons} +\rho_{fermions} = \frac{\pi^2}{30} g T^4 + 35.85 \,T^4 (q-1) ,
\end{equation}
where $g=g_b+7/8 g_f$. At  high enough temperatures, the energy
density of the universe is essentially dominated by $e^-,e^+,\nu$ and
$\hat \nu$ and so $g_b=2$ and $g_f=2+2+2 \times 3=10$.

The second correction comes from the computation of the weak interaction
rates.   We shall denote by $\lambda_{pn}(T)$ 
the rate for the weak processes to convert
protons into neutrons and by $\lambda_{np}(T)$  the rate for the 
associated, reverse ones. Within the standard approximations applicable in the
early universe regime \cite{torres1,BERNSTEIN},
it is possible to see that the weak interaction
rate $\Lambda$ is given by 
$\Lambda \simeq 4 \lambda_{\nu+n \rightarrow p+e^-}$.
This last rate for the particular reaction quoted must be computed using,

\begin{equation}
\label{r1}
\lambda_{\nu+n \rightarrow p+e^-}=A \int_0 ^\infty dp_{\nu} p_{\nu}^2
p_e E_e (1-<\hat n_e>)n_q(\nu)
\end{equation}
where $A$ is a constant fixed by the experimental value of the neutron lifetime.
Using (\ref{1},\ref{2}), we get 

\begin{equation}                 
\lambda_{\nu+n \rightarrow p+e^-}=
\lambda_{\nu+n \rightarrow p+e^-}^{standard}+
354.8 T^5 \, A (q-1), 
\end{equation}
as the biggest correction. Multiplying this result by 4, we obtain,

\begin{equation}
\frac{\delta \Lambda}{A} =  1419.2 \; T^5 \;(1-q).
\end{equation}
Having in hands the main corrections that
non-extensive quantal distribution functions provide, we move onwards
to get the final bound upon $q$. To do so, we first compute the first order 
correction to $T_f$, which is defined as the temperature where the equality

\begin{equation}
\Lambda \simeq \left( \frac{\dot a}{a} \right)= 
\sqrt{\frac{8\pi G}{3} \rho_{total}}
\end{equation} 
holds. The result is 

\begin{equation}
\frac{\delta T_f}{T_f^{st}}=6.61 (q-1).
\end{equation}
Using, finally, Eq. (\ref{b2}) we get the following bound,

\begin{equation}
\label{FF}
|q-1| < 4.01 \times 10^{-3}. 
\end{equation}

The previous bound is, as one should
expect, 
more
restrictive than that obtained within the  Tsallis et al. approach.
\footnote{Recall the Errata
in \cite{torres1}. The final bound turns out to be $|1-q| < 3.4 \times 10^{-3}$.} 
This conclude the objective of this brief letter.

Summing up, in this work we have revisited the recently proposed early
Universe test of TT and computed the related bound on $q$ using the
generalized distribution functions within the factorization approximation. 
Our main result is written in Eq. (\ref{FF}): 1 to
100 seconds after the Big Bang, $q$ must be that close 
to 1 in order to be able to reproduce observational results on helium abundance.
As it
is expected, the bound is found to be consistent with those of 
Tsallis et al. approach, a result which was also obtained in Refs.
\cite{tirnakli1,tirnakli2}
.
Although all these efforts clarify
the fact that the factorization approach results can be used with confidence
for physical systems, our belief is that it will be used much more in the
near future especially for the applications with $q$ values far from unity,
where Tsallis et al. approach is unapplicable. Therefore new attempts on
this line would be highly welcomed.\footnote{After submission of 
this work we became aware of the work by Rajagopal et al.
[Phys. Rev. Lett. 80, 3907 (1998)]  where exact results for quantal
distribution functions are given. We have explored how these results
match with Tsallis et al.'s and factorization approaches and we shall
report on it in a future comunication.}

\section*{Acknowledgments}

U.T. is a  TUBITAK M\"{u}nir Birsel Foundation Fellow and acknowledges
partial support from Ege University Research Fund under the Project Number
97 FEN 025. During the course of this research, 
D.F.T. was a Chevening
Scholar of the British Council Foundation 
and acknowledges partial support from
CONICET and Fundaci\'on Antorchas.

\end{document}